\documentclass[sigconf]{acmart}

\usepackage{booktabs} 

 \renewcommand\footnotetextcopyrightpermission[1]{} 
\usepackage{mathtools}

\usepackage{arydshln,leftidx}
\usepackage{amsmath}
\usepackage{amssymb}
\usepackage{kbordermatrix}
\usepackage{graphics}
\usepackage{graphicx}
\usepackage{enumitem}
\usepackage{textcomp}
\usepackage{color}
\usepackage{outlines}

\usepackage{amsthm}

\usepackage{caption}
\usepackage{setspace}
\usepackage{threeparttable}

\usepackage{changepage}   

\usepackage{etoolbox}
\let\bbordermatrix\bordermatrix
\patchcmd{\bbordermatrix}{8.75}{4.75}{}{}
\patchcmd{\bbordermatrix}{\left(}{\left|}{}{}
\patchcmd{\bbordermatrix}{\right)}{\right|}{}{}

\usepackage{blkarray}
\usepackage{multirow}
\usepackage{wrapfig}
\usepackage{tikz}

\usepackage{subfig}
\usetikzlibrary{matrix,fit}

\newcommand*\circled[1]{\tikz[baseline=(char.base)]{
            \node[shape=circle,draw,inner sep=0.3pt] (char) {#1};}}

\begin{document}
\title{BRISC-V Emulator: A Standalone, Installation-Free, Browser-Based Teaching Tool}


\author{Mihailo Isakov and Michel A. Kinsy}
\affiliation{%
 Adaptive and Secure Computing Systems (ASCS) Laboratory \\
Department of Electrical and Computer Engineering \\
  \institution{Boston University}
}
\email{{mihailo, mkinsy}@bu.edu}
%
%


\begin{abstract}
 Many computer organization and computer architecture classes have recently started adopting the RISC-V architecture as an alternative to proprietary RISC ISAs and architectures. Emulators are a common teaching tool used to introduce students to writing assembly. We present the BRISC-V (\underline{B}oston University \underline{RISC-V}) Emulator and teaching tool, a RISC-V emulator inspired by existing RISC and CISC emulators. The emulator is a web-based, pure javascript implementation meant to simplify deployment, as it does not require maintaining support for different operating systems or any installation. Here we present the workings, usage, and extensibility of the BRISC-V emulator. 
\end{abstract}

\keywords{RISC-V, ISA, assembly, simulation, emulation, web-based, teaching, tool.}

\maketitle

\section{Introduction}
The x86 architecture assembly language is typically considered too large and complex to be taught in a college entry-level class on assembly or computer organization. Instead, RISC architectures are a common choice due to their relatively small instruction set.
A student learn to write assembly by hand using RISC emulators either through the command line, or wrapped in a graphical user interface (GUI). The student uses the emulator by first loading an assembly program written in an external editor, which is then wrapped in a small kernel. The kernel is tasked with initializing the registers and providing a small number of system calls. The user then either runs or steps through the program line-by-line. The emulator offers the student a view of the code with the instruction pointer clearly visible, as well as tables showing the state of the registers and memory when the program is paused or has finished. 

RISC-V is an open-source RISC architecture widely seen as an alternative to proprietary RISC and CISC architectures. It has seen broad adoption in the recent years, with a number of extension proposals, public implementations, growing software support, several fabricated designs, and software emulators. It is increasingly being used in college-level classes as a more modern RISC alternative, but the teaching tools have yet to catch up.

In this paper, we present our browser-based RISC-V emulator we call BRISC-V Emulator (\underline{B}oston University \underline{RISC-V} \underline{Emulator}). It is inspired by existing RISC-based emulators, and has been developed for the purpose of teaching computer organization and computer architecture classes. 

Due to common issues with (1) installing software on locked-down machines and (2) students wanting to use their own machines with different operating systems (OSs), the authors wanted a javascript-based, OS-oblivious RISC-V emulator that does not require installation by the user, and provides a consistent GUI independent of the platform. Having a browser-based implementation reduces the need to maintain multiple versions of the codebase, and removes the perceived `bloat' from installing new software. BRISC-V emulator was built with simplicity and extensibility in mind.

\section{BRISC-V Emulator}
The whole emulator is written in javascript, and runs on the user's computer. The emulator can be ran in three ways: (1) the student can download it and run it, i.e. by unzipping a directory and clicking on a \texttt{index.html} file which opens in the browser, (2) hosted by a teaching assistant which would give out students a link to access the emulator, or (3) from our website, located at 

\textbf{{\small \texttt{http://ascslab.org/research/briscv/emulator/emulator.html}}}.

After the page is loaded, all further processing is done on the students computer. This is beneficial when hosting the emulator for a large class, as the server can cache the website, and performs no other processing once the site is loaded.

\begin{figure}[h]
\begin{center}
    \includegraphics[width=1.0\columnwidth]{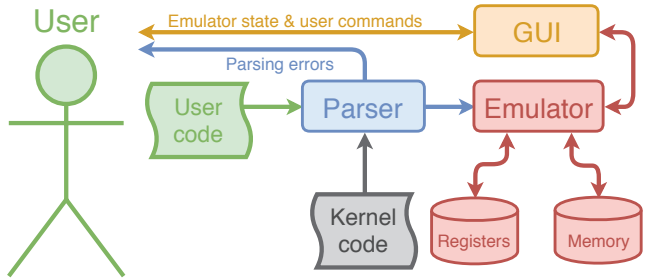}
    \caption{The emulator pipeline.}
    \label{fig:pipeline}
\end{center}
\end{figure}

The emulator is roughly split into three parts: the parser, the emulator, and the GUI, as seen in Figure~\ref{fig:pipeline}. The user writes RISC-V assembly code in an external text editor, and loads the code through the website. The parser first wraps the code with the kernel code, which consists of the front-end code setting register states as well as the stack pointer, and backend code which contains an infinite loop used in place of an exit system call. The parser then parses the code and produces a list of instruction objects and a label map. Each element in the instruction list is an object containing a list of symbols, the instruction type (instruction, label, pseudo-instruction, directive, or error), the line number in the original file, and a instruction address. The label map maps labels to appropriate instruction addresses.

The emulator accepts the instruction list and label map, and initializes an array of registers and a memory map. The registers are small enough and frequently used enough that we create a simple array to store them. We use a map for the memory as most programs will not allocate large amounts of memory. When a certain address is accessed, if not already present, it is created in the map, where the key is the address and the value is provided by the program.
The emulator provides several functions tied to GUI buttons: (1) stepping through one instruction, (2) running the code until the backend infinite loop or a breakpoint is hit, and (3) initializing the emulator. The \texttt{single\_step} function takes the instruction at the current instruction pointer (IP) and depending on the type of instruction, updates the registers, memory, and IP. How each instruction is handled (\texttt{ADD}, \texttt{JALR}, \texttt{AUIPC}, etc.) is hardcoded in the emulator. The \texttt{run} instruction repeatedly calls the \texttt{single\_step} instruction until it detects that the PC has hit the backend kernel infinite loop or breakpoint. The \texttt{setup\_emulator} is used to initialize or reset the emulator. It calls the appropriate parser functions, stores parsers outputs, and initializes the registers and memory. 

\section{Using the BRISC-V Emulator}
\begin{figure*}[h!]
\begin{center}
   \includegraphics[width=\textwidth]{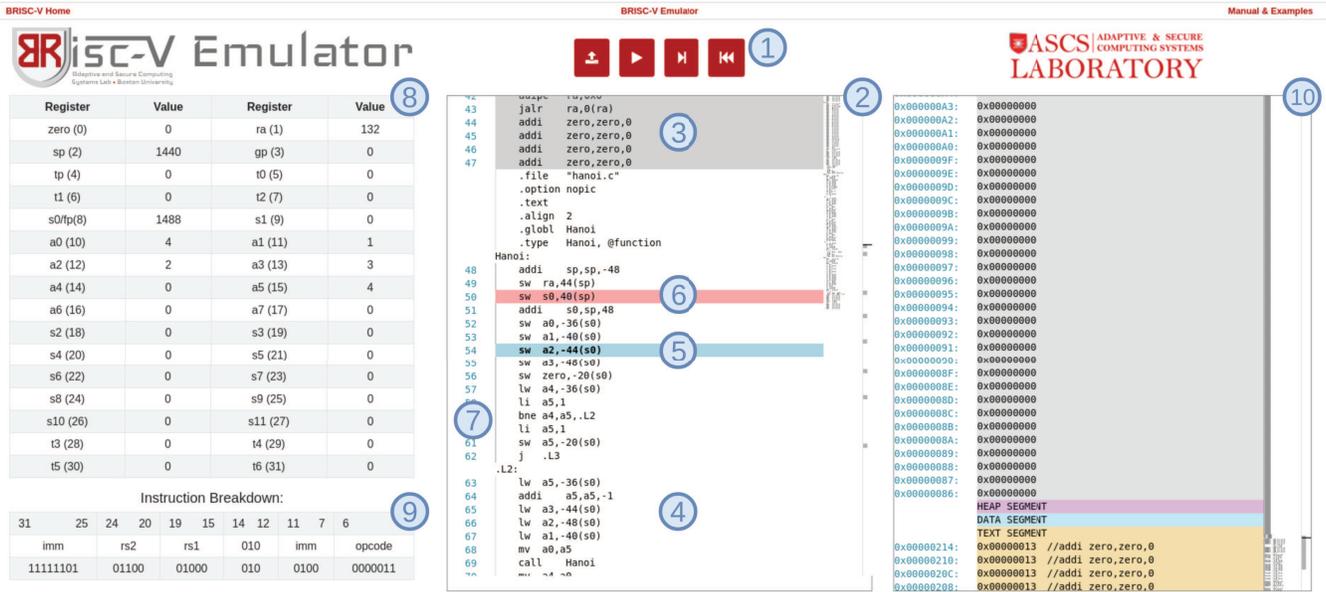}
    \caption{A screenshot of the BRISC-V emulator user interface. }
    \label{fig:screenshot}
\end{center}
\end{figure*}

Figure~\ref{fig:screenshot} shows a screenshot of the current version of the emulator web-page. The page is split into three columns. In the middle column, we have buttons for loading assembly, running the loaded code, stepping through one instruction, and restarting the emulator \circled{1}. Below the buttons is a code editor \circled{2} (though currently read-only). The editor shows both the kernel lines \circled{3} (with a grey background), as well as user's code \circled{4} (with a white background). The current instruction is highlighted blue \circled{5}. Each instruction has an instruction pointer before it \circled{6} (shown here in decimal). Right-clicking on an instruction opens a context menu where the user can set or remove a breakpoint at the selected instruction. When running the code (not stepping through it), the run will stop when hitting a breakpoint. This allows the user to skip large parts of assembly and quickly debug regions of interest. The editor allows code-folding, where the user can hover left of any label and a \textbf{+} symbol appears. Clicking the symbol folds/unfolds the instructions between the specified and the next label.

On the left-hand column, the user can monitor registers \circled{7} and view the instruction at the current IP in binary \circled{8}. The register file displays the values of all 32 registers in decimal, hexadecimal or binary representations. As the IP progresses through the code, some registers will change values, and this change will be highlighted by changing the background of the appropriate register to red. At the same time, each new (non-pseudo) instruction will be broken down to its 32bit instruction format, shown on the bottom-left side of Figure~\ref{fig:screenshot} \circled{8}. The elements in the top row of this table signify the bitrange of the column. The middle row specifies which part of the instruction is shown in the column, and changes with instruction type, i.e., R-type instructions will have columns \texttt{funct7}, \texttt{rs2}, \texttt{rs1}, \texttt{funct3}, \texttt{rd}, \texttt{opcode}, while I-type instructions will have one less column, as \texttt{funct7} and \texttt{rs2} are replaced with \texttt{imm}.

\begin{figure}[H]
\begin{center}
    \includegraphics[width=\columnwidth]{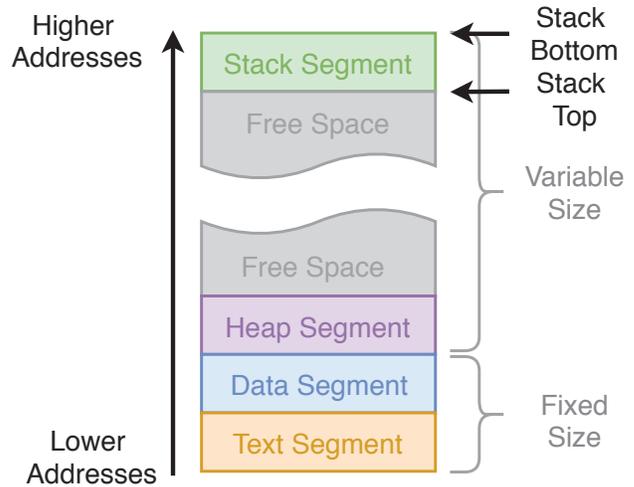}
    \caption{Emulator program memory management.}
    \label{fig:memory}
\end{center}
\end{figure}

On the right column is the memory visualization \circled{9} table. It is a descending list with 5 foldable regions: the stack segment, free space, heap segment, data segment, and text segment. Each line represents one word (32 bits). The address is shown on the left in hexadecimal format (light blue), and the value is shown on the right. Lines belonging to the text segment also contain a comment specifying which instruction is stored at that location. Same as with labels in the code (middle) column, each of the segments can be folded. Figure~\ref{fig:memory} depicts the memory management layout.

\section{Extensibility}
While there exist more mature RISC-V emulators like the RISC-V Angel~\cite{angel}, we choose to write our own for multiple reasons: (1) RISC-V Angel seems focused more on performance (booting BusyBox in seconds) than on showing the state of the processor at every instruction, (2) RISC-V Angel is harder for students to modify. We aim to provide students with a simple, `hackable' tool which they can use in computer organization classes to get familiar with assembly, in computer architecture classes to test out existing ISA extensions (i.e. floating point or vector instructions) and confirm their compiled code behaves as expected, and in hardware security classes to quickly test out new security ISA extensions.

As an example, to implement the multiplication operations \texttt{MUL}, \texttt{MULH}, \texttt{MULHU}, \texttt{MULHSU} proposed in the RISC-V specification~\cite{spec}, one needs to: 
\begin{enumerate}
\item Add the \texttt{MUL}, \texttt{MULH}, \texttt{MULHU}, \texttt{MULHSU} to the \texttt{instructions.js} file containing instruction names.
\item Edit the \texttt{parser.js} file so that these instructions are parsed with the two source registers and a destination register. This requires minimal coding as the registers are already extracted by the parser.
\item Add a new case condition in the \texttt{emulator.js} that on a \texttt{MUL*} instruction (1) updates the IP by 4, and (2) puts the correct multiplication result in the correct register. The emulator already has the registers as local variables, so the user just needs to refer to the appropriate ones.
\end{enumerate}

\section{Conclusion}
We introduced in this brief the BRISC-V emulator that targets the RISC-V architecture. The emulator is meant to be used as a teaching tool, as currently many computer organization and computer architecture classes are migrating to RISC-V. The emulator is written in vanilla javascript, and the user only needs to open an HTML page to use it. It does not require any installation, simplifying class setup as no admin access is needed on student machines. The emulator will be made available at the following address upon publication: \textbf{{\small \texttt{http://ascslab.org/research/briscv/emulator/emulator.html}}}.

\bibliographystyle{ACM-Reference-Format}
\bibliography{paper} 

\end{document}